\documentclass[aps,prb,floatfix,twocolumn,showpacs,preprintnumbers,amsmath,amssymb,superscriptaddress]{revtex4}

\usepackage{graphicx}  
\usepackage{dcolumn}   
\usepackage{bm}        
\usepackage{stmaryrd}  
\usepackage{multirow}  
\usepackage{color}     
\usepackage{subfigure}

\def   \Ms       {M_{\rm S}}                             
\def   \Msfixed  {M_{\rm S\, fixed}}                        
\def   \lexch    {l_{\rm exch}}                             
\def   \heff     {{\textbf{h}}_{\rm eff}}             
\def   \Ku       {K_{\rm u}}                             
\def   \mi       {\mu_{\rm 0}}
\def   \tor      {\bm \tau}                              
\def   \tin      {{\bm \tau_{\theta}}}                   
\def   \tout     {{\bm \tau_{\varphi}}}                       
\def   \nm       {{\rm nm}}                              

\def   \gama     {{\gamma_{0}}}

\def   \bfm      {{\bf m}}

\def   \bfp      {{\bf p}}

\begin{document}

\title{Computational study of microwave oscillations in absence of external field in nonstandard spin valves in the diffusive transport limit}

\author{E.~Jaromirska}
\affiliation{Departamento de F\'isica Aplicada, Universidad de Salamanca, Spain}
\author{P.~Bal\'a\v{z}}
\affiliation{Department of Physics, Adam Mickiewicz University,
             Umultowska 85, 61-614~Pozna\'n, Poland}
\author{L.~L\'opez D\'iaz}
\affiliation{Departamento de F\'isica Aplicada, Universidad de Salamanca, Spain}
\author{J.~Barna\'s}
\affiliation{Department of Physics, Adam Mickiewicz University,
             Umultowska 85, 61-614~Pozna\'n, Poland}
\affiliation{Institute of Molecular Physics, Polish Academy of Sciences
              Smoluchowskiego 17, 60-179 Pozna\'n, Poland}

\date{\today}

\begin{abstract}
An anomalous (inverse) spin accumulation in the nonmagnetic spacer
may build up when the spin valve consists of magnetic films having
different spin asymmetries. This leads to {\em wavy}-like
dependence of spin-transfer torque on the angle between
magnetizations, as predicted by spin-dependent diffusive transport
model, and also confirmed experimentally. Making use of these
predictions, we have numerically studied the magnetization
dynamics in presence of such a {\em wavy} torque in  Co(8
nm)/Cu(10 nm)/Py(8 nm) nanopillar, considering geometry with
extended and etched $Co$ layer. In both cases we specify
conditions for the out-of-plane precession to appear in absence of
external magnetic field and neglecting thermal fluctuations. We
prove the assumption of {\em wavy}-like torque angular dependence
to be fully consistent with experimental observations. We also
show that some features reported experimentally, like nonlinear
slope of frequency vs. current behavior, are beyond the
applicability range of macrospin approximation and can be
explained only by means of full micromagnetic analysis.
\pacs{67.30.hj,75.60.Jk,75.70.Cn,78.20.Bh}
\end{abstract}

\maketitle

\section{Introduction}\label {Sec:introduction}
\typeout{--------------------------------------}
\typeout{Introduction}
\typeout{--------------------------------------} The concept of
spin transfer was introduced in pioneering works by
Slonczewski\cite{Slonczewski1996:JMMM} and
Berger\cite{Berger1996:PRB}. They have shown that spin polarized
current can exert torque on a thin magnetic film due to transfer
of spin angular momentum, influencing the magnetic state of the
layer. The spin-transfer torque (STT) originates from spin
asymmetries of the two independent transport channels, and its
well known manifestation is current induced magnetic switching
(CIMS)\cite{AlHajDarwish:PRL,Tsoi2004:PRB} as well as generation
of microwave oscillations\cite{Slavin:PRB,Kiselev2003:Nature}.
Properties of STT are related to the sample design and material
parameters. This also implies that CIMS and
current-perpendicular-to-plane giant magnetoresistance (CPP-GMR)
phenomena are correlated\cite{Gmitra2009:PRB} and depend on the
same structural parameters. By considering two well-defined
conduction spin-channels, Valet and Fert\cite{Valet:PRB}
incorporated most of these parameters into CPP-GMR model based on
spin diffusion transport equations. Generalization of Valet-Fert
approach\cite{Barnas2005:PRB} includes STT and provides an unified
description of STT and CPP-GMR in the diffusive transport limit.
For symmetric spin valves, with fixed and free layer made of the
same material, such as Co/Cu/Co, this model predicts standard
behavior of STT, which does not vary qualitatively from
Slonczewski's result obtained in the ballistic transport
limit\cite{Slonczewski1996:JMMM}. In this case, current of one
orientation drives switching to antiparallel configuration while
opposite current stabilizes parallel state\cite{Katine:PRL}. In
presence of applied fields higher than the coercive field, the
generation of microwave oscillations is
possible\cite{Kiselev2003:Nature}. Similar behavior has been
recently observed also in Py/Cu/Py (Py = Permalloy), both
experimentally\cite{KrivorotovScience:Science,KrivorotovPRB2008:PRB}
and theoretically \cite{KrivorotovPRB2007:PRB}. Moreover, standard
STT associated precession was also reported at high perpendicular
magnetic field in an asymmetric spin valve\cite{Kiselev2004:PRL},
where the magnetic layers are made of different materials (Co(40
nm)/Cu(10 nm)/Py(3 nm)).

However in such asymmetric structures, provided that spin
asymmetries and spin diffusion lengths differ markedly and the
thickness obeys certain conditions, qualitatively different
situation may arise. The STT vanishes and changes sign in a
certain noncollinear magnetic configuration ({\em wavy}-like STT)
due to the appearance of an inverse spin accumulation in the
nonmagnetic spacer. Therefore, current flowing in one direction
destabilizes both collinear magnetic configurations, whereas the
opposite current stabilizes both of them. The first case is of
particular interest as it leads to excitation of stationary
oscillation modes in absence of external magnetic field.

In this paper we present a systematic study of dynamic response of
a magnetic film to such {\em wavy}-like STT. The asymmetric pillar
under study is a spin valve consisting of a fixed layer,
nonmagnetic spacer and a free magnetic layer, Co(8 nm)/Cu(10
nm)/Py(8 nm) respectively, with the elliptical cross-section of
$155\times100$ $nm^{2}$. The polarization of the fixed
layer is assumed to be along the ellipse major axis. Recently,
such asymmetric structures have been investigated
theoretically\cite{Barnas2005:PRB,Gmitra2006:APL,Gmitra2007:PRL}
as well as experimentally at low and zero applied
fields\cite{Boulle2007:NP,Boulle2008:PRB}. To authors' best knowledge, no micromagnetic analysis of such
structures in the diffusive transport limit has been performed so
far. Moreover, the reported macrospin (referred to in the
following also as single domain) study does not describe correctly
the dynamics at low applied magnetic fields\cite{Boulle2008:PRB}.
Here, starting with the single domain approximation, and extending
study to the full micromagnetic model, we explain the origin of
out-of-plane precession (OPP). In particular, we show that only
full micromagnetic model can successfully reproduce magnetization
dynamics at low applied magnetic field.

The paper is organized as follows. Section \ref{Sec:torque}
describes briefly the torque calculations in the diffusive
transport limit. The methodology of simulations is presented in
section \ref{Sec:simulation}. The results of numerical study and
their discussion are to be found in section \ref{Sec:results}, whereas final
conclusions in section \ref{Sec:conclusion}.

\section{Spin torque in an asymmetric pillar}\label{Sec:torque}
\begin{figure}[!t]
  \subfigure[]{\includegraphics[width=0.48\columnwidth]{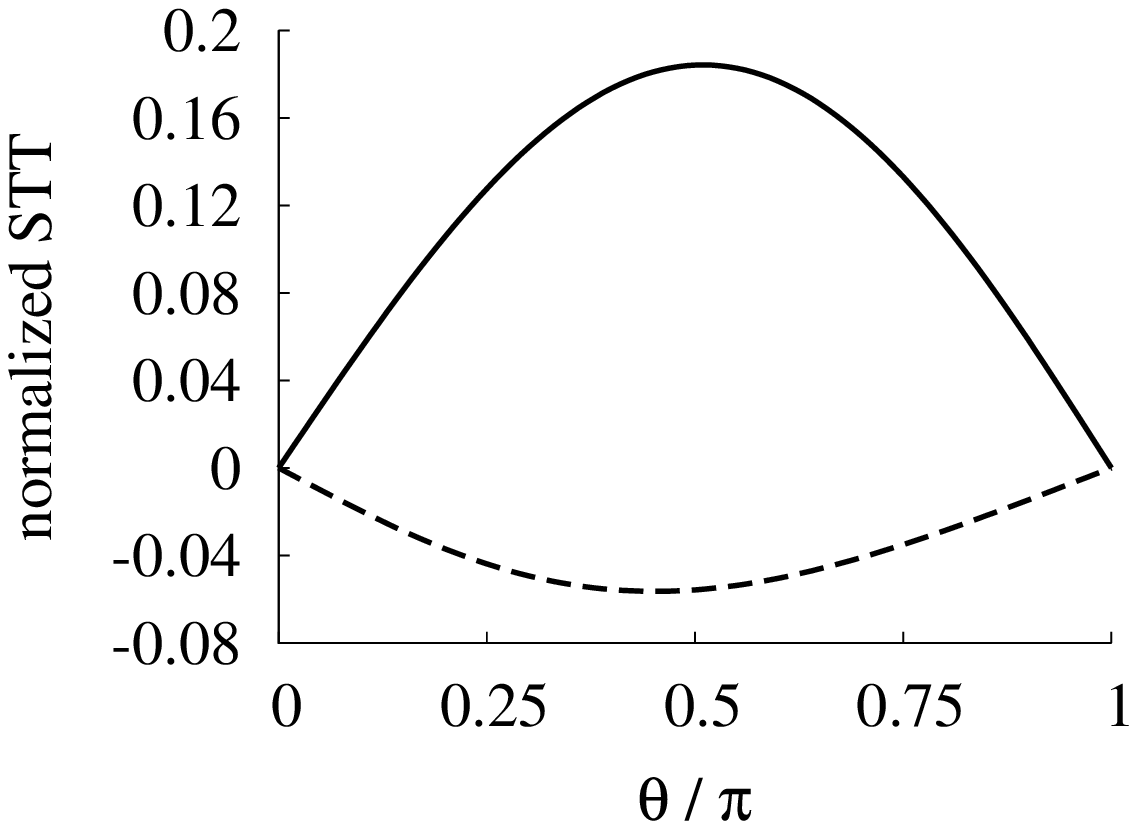}}
  \hfil
  \subfigure[]{\includegraphics[width=0.48\columnwidth]{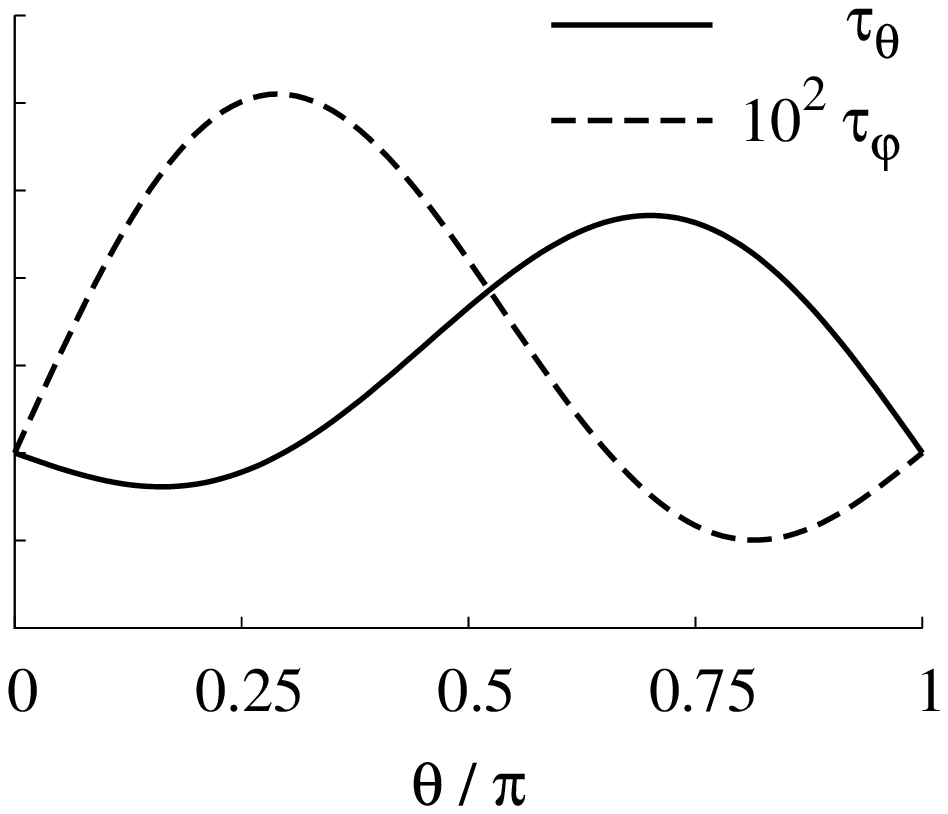}}
  \caption{Angular dependence of the spin transfer torque normalized to $\hbar
  j/|e|$ for
           (a) symmetric Py(8nm)/Cu(10nm)/Py(8nm) and
           (b) asymmetric Co(8nm)/Cu(10nm)/Py(8nm) pillar.
           }
\label{Fig:wavy_torque}
\end{figure}

\typeout{--------------------------------------} \typeout{Spin
torque calculations}
\typeout{--------------------------------------} As mentioned in
the introduction, structure of a pillar determines the dependence
of  STT on the angle between magnetization vectors. Generally, STT
consists of two components, $\tor = \tin  + \tout$, where  $\tin$
is the in-plane (IP) component, while
 $\tout$ is the out-of-plane (OP) one.  These two components can be written
 as\cite{Barnas2005:PRB}
\begin{subequations}
\label{Eq:torques}
  \begin{align}
  \label{Eq:torqueIP}
    \tin &= -aj\,\bfm \times ( {\bf m} \times {\bf p})\, , \\
  \label{Eq:torqueOP}
    \tout &= bj\,{\bf m} \times {\bf p}\, ,
  \end{align}
\end{subequations}
where ${\bf m}$ denotes the normalized (unit) vector along the
magnetization of the free layer, ${\bf p}$ is the normalized
magnetization of the fixed layer, and $j$ is the current density.
The prefactors $a$ and $b$ are independent of current $j$,  but
they generally depend on the angle $\theta$ between ${\bf m}$
and ${\bf p}$. These parameters have been computed from the
diffusive transport model\cite{Barnas2005:PRB}. First, from the
boundary conditions  for spin current and spin
accumulation\cite{Brataas:EPJ} we calculate the spin current. The
torque is then calculated from the normal component of the spin
current in the nonmagnetic film at the interface with the magnetic
layer. Most of the parameters used in this description, like
interface and bulk spin asymmetry coefficients, interface
resistances, layer resistivities\cite{BassJMMM:JMMM}, and spin
diffusion lengths\cite{BassJPCM:JPCM} are provided by the
corresponding CPP-GMR experiments. The two remaining parameters,
i.e. the real and imaginary part of the mixing
conductance\cite{Brataas2006:PR} have been extracted from spin
current interface transmission calculations\cite{Stiles:JAP}. The
variation of normalized (to $\hbar j/|e|$) STT with the angle
$\theta$ for symmetric and asymmetric spin valve is shown in
Fig.~\ref{Fig:wavy_torque}.

In the system under considerations (Fig.~\ref{Fig:wavy_torque}),
the OP torque is roughly 2 orders of magnitude smaller than the IP
component, which means that the latter will not markedly influence
the magnetization dynamics. Comparing the angular dependence for
the standard (Fig.~\ref{Fig:wavy_torque}a) pillar structure with
that for the nonstandard one (Fig.~\ref{Fig:wavy_torque}b), the
uniqueness of the latter is clearly visible -- at some critical
angle $\theta$ the IP component of the torque vanishes. This gives
rise to interesting dynamics at zero and low magnetic fields.
Above certain threshold current both collinear states of the
magnetization are unstable for one current orientation, and the
only solution of the Landau-Lifshitz-Gilbert (LLG) equation is
then the steady state precession or a noncollinear static
magnetization state\cite{Barnas2006:MSEB}.

\section{Simulation methodology}\label{Sec:simulation}
\typeout{--------------------------------------}
\typeout{Simulation methodology}
\typeout{-------------------------------------}

Magnetic dynamics of the free layer is described by the
Landau-Lifshitz-Gilbert (LLG) equation with the STT included. This
equation, when written as the equation for time evolution of the
unit vector $\bfm$, takes the form
\begin{center}
 \begin{eqnarray}
\label{Eq:LLG}
  \frac{1+\alpha^{2}}{\gama\Ms}\, \frac{d \bfm}{dt} =
  -\bfm \times \heff -
  \alpha  \bfm \times (\bfm \times \heff) \nonumber \\
  +   \frac{j}{\mu_0 \Ms^2 d}\, [\tilde{a}\bfm\times(\bfm\times\bfp )-\tilde{b}\bfm\times\bfp ]
\end{eqnarray}
\end{center}
where $\tilde{a}=a-\alpha b$, $\tilde{b}=b+\alpha a$, $\heff$ is
the effective field normalized to the saturation magnetization
$\Ms$ of the free layer, $\alpha$ is the damping constant, $d$ is
the thickness of the free layer, and $\gama =|\gamma| \mu_0$ with
$\gamma$ being the gyromagnetic ratio and $\mu_0$ standing for the
magnetic vacuum permeability. Since $\alpha$ is a small parameter,
in numerical simulations we assumed $\tilde{a}=a$ and
$\tilde{b}=b$.

In our convention the current flowing from the fixed towards the
thin (free) layer defines the positive current direction, $j
> 0$. Note, that in this convention the positive current triggers
switching to the antiparallel configuration, whereas negative
current stabilizes parallel alignment in standard spin valves. On
the other hand, negative current in nonstandard spin valves
supports oscillatory regime, while positive current stabilizes
both collinear configurations.

Only the dynamics of the free layer is resolved, while
magnetization of the fixed layer is assumed to remain uniform in
the film plane. Apart for this, we have neglected all thermal
effects. In the macrospin analysis, the effective field is assumed
to include the self-magnetostatic term\cite{Aharoni:JAP}, the
uniaxial anisotropy field, and the external magnetic field. The
uniform magnetostatic coupling field (with the fixed layer) can
also be taken into account. In the micromagnetic study, on the
other hand, the magnetic field is modified with respect to
macrospin case. First, the self-magnetostatic term is computed
using the Fast Fourier Transform technique assuming that the
magnetization is uniform in each computational
cell\cite{Newell:JGR}. Second, six-neighbor dot product
representation is used to compute the exchange field. Unless
stated differently in the text, the following values of the
relevant parameters have been chosen for the systematic study: the
anisotropy constant $\Ku=3.46\times 10^{3}$ $J/m^{3}$, exchange
constant $A=1.3\times 10^{-11}$ $J/m$, damping constant
$\alpha=0.01$, $\Ms =6.9\times10^{5}$ $A/m$, and  $\Msfixed
=1.4\times 10^{6}$ $A/m$. As follows from this choice for permalloy,
the corresponding exchange length $\lexch = \sqrt{ 2A/\mi
\Ms^{2}}$ equals $6.6$ $\nm$, and therefore we choose
$5\times5\times4$ $\nm^{3}$ discretization mesh. Further refinement
of the cell size in the $z$-direction does not lead to any
substantial difference in system frequency response. Fourth-order
Runge-Kutta scheme was employed for the time integration of
Eq.~\ref{Eq:LLG}, and the stability analysis was carried out for
the chosen mesh to assign appropriate time integration step.

\section{Results and discussion}\label{Sec:results}
\typeout{--------------------------------------} \typeout{Results
and discussion} \typeout{--------------------------------------}

Before comparing results obtained from macrospin approximation
with those from full micromagnetic study, one should consider that
the concept of single domain magnetic particle in many situations
is not justified. We shall explain how the difference between the
results of macrospin and micromagnetic analysis arises from the
appearance of an inhomogeneous magnetization and a finite exchange
field. The exchange energy density of a closed $\bm{M}(\bm{r})$
configuration increases as the particle size decreases, and this
could, in principle, justify macrospin approach below certain
critical size of the system, even though spin torque makes the
estimation of this critical size difficult\cite{BerkovJMMM:JMMM}.
However, as reported in [\onlinecite{BerkovPRB71:PRB}], steady
state precession of a thin square nanoelement exhibits complicated
transition from quasi-macrospin to chaotic behavior already at the
size of 30 nm, which invalidates single domain approximation for
most of experimentally studied systems. Moreover, on the basis of
micromagnetic analysis Berkov and Gorn\cite{BerkovPRB72:PRB}
identified some artifacts of macrospin model in the ballistic
transport limit. These artifacts might cause misleading
interpretation of the origin of some observed
phenomena\cite{Kiselev2003:Nature}. Aware of that, we cannot
however deny that macrospin dynamics is very rich and in many
cases can serve as a tool for the basic understanding of physics
behind.

Since the precessional states appear for negative current
(according to the definition introduced in section III), we limit
the following considerations to negative current. For sake of simplicity, the current $I$ denotes the absolute value of the
negative current.

\subsection{Extended geometry}

\begin{center}
  \begin{figure}[h]
\subfigure[]{\includegraphics[width=0.48\columnwidth]{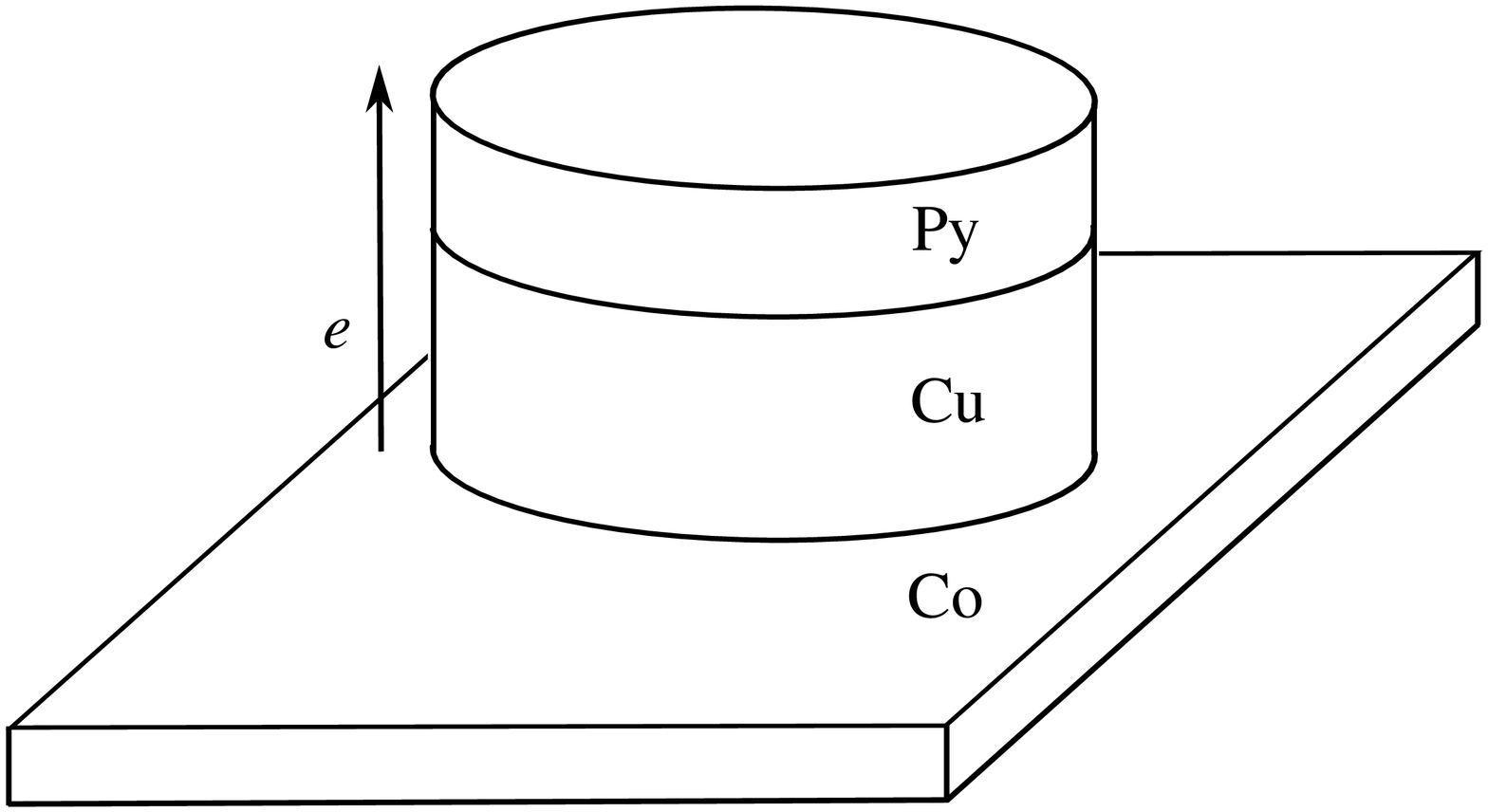}
}
\hspace{20pt}
\subfigure[]{\includegraphics[width=0.22\columnwidth]{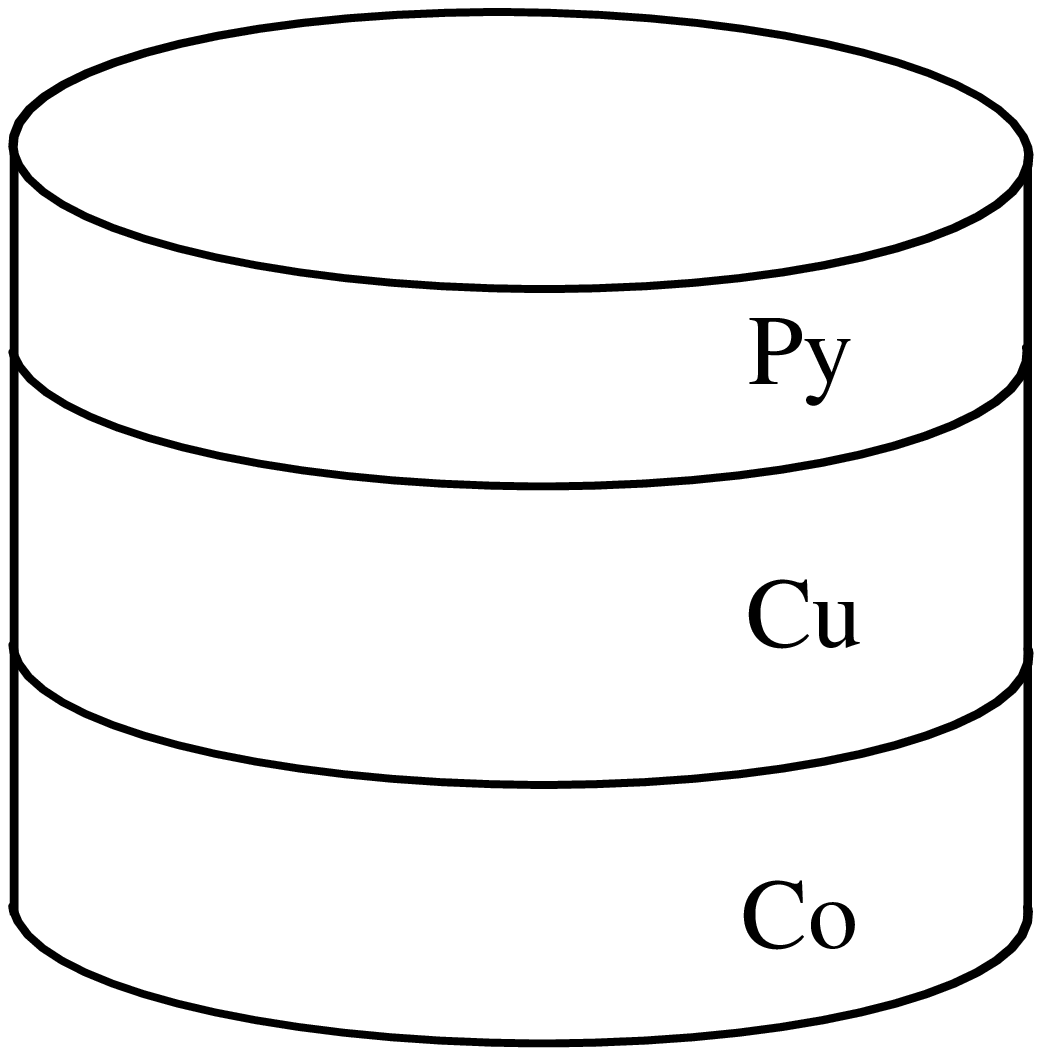}
}
    \caption{Geometry of a pillar with (a) an extended and (b) etched fixed layer.
    The interlayer coupling field in (a) can be neglected.}
  \label{Fig:extended}
  \end{figure}
\end{center}

We start our numerical study with a pillar structure having an
extended fixed layer, like the one shown schematically in
Fig.~\ref{Fig:extended}a. In this case, the influence of
interlayer magnetostatic coupling field (ICF) can be neglected.
Since the anisotropy of Py is also small, one concludes that it
is mainly the self-magnetostatic field that drives the system
dynamics. In the following we consider the role of initial
magnetic state of the system, and begin with the parallel magnetic
configuration.

\subsubsection{Initial P state}

Assume the system is initially in the parallel configuration and
increase the current step-like (with the step $\Delta I$) in order
to identify possible oscillatory regimes. As the initial state for
simulation at a given current $I$ we assume the state the systems
arrived at in the preceding step of simulations, i.e. for current
$I-\Delta I$. Simulation results from both macrospin approximation
and micromagnetic model reveal that an in-plane precession (IPP)
is supported at zero applied field, which is consistent with
previous study [\onlinecite{Gmitra2006:APL, Gmitra2007:PRL}]. The
associated frequency redshift with current is a well
known phenomenon and is attributed to the increase of the
oscillation amplitude. Interestingly the redshift in the
micromagnetic model turns out to be much slower than in the
macrospin approximation, which is opposite to the results obtained
in the ballistic transport limit\cite{BerkovJMMM:JMMM}. We shall
explain this effect later on.

\begin{center}
  \begin{figure}[h]
    \includegraphics[angle=0,width=1\columnwidth]{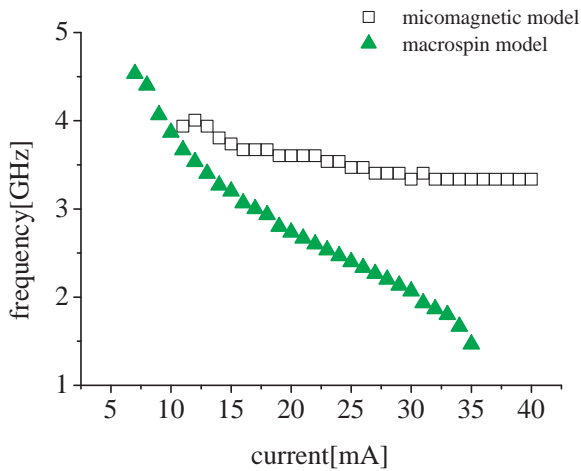}
    \caption{Frequency vs. current behavior in absence of external field in the macrospin
            model - full triangles, and in micromagnetic model - open squares.
            This convention is kept throughout this paper:
            open symbols correspond to results of micromagnetic study and
            full symbols to those in the macrospin model.
            A 'cut-off' current is observed at I=35 mA in the single domain model.
            The simulations have been performed for the step-wise increasing
            current, and for
            parallel initial configuration of the system. Initial
            state for a current $I+\Delta I$ was taken as the
            state at which the system arrived at in the simulations for the current $I$.
            }
 \label{Fig:zero_field}
  \end{figure}
\end{center}

\begin{center}
  \begin{figure}[h]

    \includegraphics[angle=0,width=0.7\columnwidth]{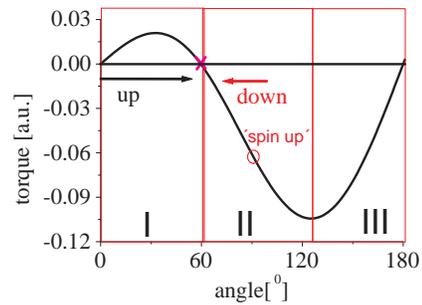}
    \caption{The torque normalized to $\hbar j/|e|$ for negative current density
             ($j<0$, compare Fig.1)
             acting on the free layer.
             As the current is increased, the frequency redshift due to the amplitude
             increase is observed
             (angle increase, region I). When the critical angle is reached
             (cross between the regions I and II),
             the torque vanishes and the magnetization tends to align along the effective
             field driving system out
             of the critical angle towards the stable static 'spin-up' state.}
  \label{Fig:explanation}
  \end{figure}
\end{center}

As follows from Fig.~\ref{Fig:zero_field}, the precessional states
in the macrospin model disappear above a certain 'cut-off' current
(I=35 mA), where a stable static 'spin-up' (normal to the film
plane) magnetization state is formed. Existence of this static
state has been recently reported in [\onlinecite{Gmitra2006:APL,
Balaz2009:PRB, Balaz2008:APP}], where it was found that the LLG
equation has just two possible solutions -- self-sustained
precession or a stable static state (above the 'cut-off' current).
The appearance of the latter can be explained with the help of
Fig.~\ref{Fig:explanation}. As the current is applied and the
system is initially in the P state, the STT counterbalances the
damping, and steady state $\rm IPP_{+}$ is obtained in region
I. When the current is increased, the oscillation amplitude
increases and therefore the critical angle (marked as cross), at
which the torque vanishes, is reached. The magnetization starts
then aligning along the effective magnetic field, and finally the
stable 'spin-up' state is reached. No such static state has been
reported in micromagnetic simulations. Assume now that the system
in macrospin simulations is in the static 'spin-up' state and then
the current is decreased (region II). The self-magnetostatic field
in the static point together with the STT may trigger the OPP in
some current range, as shown in Ref.
[\onlinecite{Gmitra2007:PRL}]. Again, no OPP appears in the
micromagnetic simulations. In general, when the angle between
magnetizations of the free and fixed layer is lower than the
critical angle (region I), only IPP is observed. As the critical
angle is crossed (region II) the OPP can be triggered in the
macrospin model, as shown in Refs. [\onlinecite{Jaromirska:tbp,
Gmitra2006:APL}]. However, we point again that the 'spin-up' state
is not reached in micromagnetic simulations because of the
inhomogeneous character of the magnetization. Therefore, even
though single domain model predicts OPP, which was also reported
experimentally\cite{Boulle2008:PRB}, the interpretation of its
origin requires further considerations.

\subsubsection{Initial AP state}

As we have seen above, two factors play crucial role in the
excitation of precessional states in the system. These are the
magnetization state which imposes the initial self-magnetostatic
field, and the initial angle between the magnetic layers, which
determines the initial torque sign and strength. The IPP is
supported in region I (Fig.~\ref{Fig:explanation}), whereas in
region II the static state has been found in the macrospin
approximation so far.

\begin{figure}[h]
\centering
\includegraphics[angle=0,width=1\columnwidth]{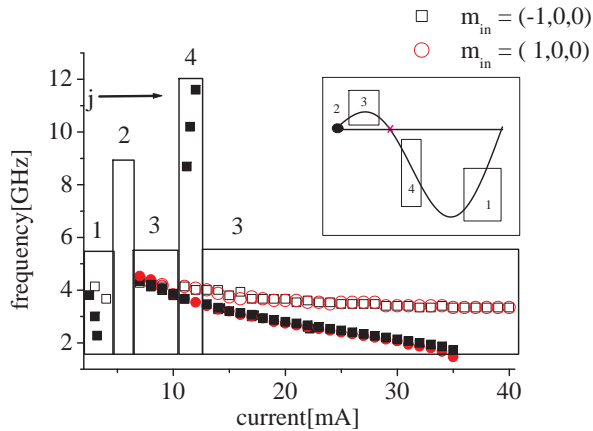}
\caption{The influence of the initial magnetization state on the dynamic
behavior of the system and the
corresponding spots in torque diagram (inset). Circles refer to P, whereas squares to AP initial state. At low currents, $\rm IPP_{-}$ mode
(close to the AP direction) is found (marked as 1). As the
current increases, the angle of the orbit increases as well, and
$\langle m_{x}\rangle $ switches towards the P state. No
self-sustained oscillations are observed in the area 2. Further
current increase leads to the onset of the second red-shifting
branch, $\rm IPP_{+}$, marked with 3. Additionally at a certain
threshold current, the combined effect of self-magnetostatic
field of the AP initial state and negative torque sign can trigger
OPP in the macrospin model, marked as 4.} \label{Fig:AP_both}

\end{figure}

In order to investigate modes supported in region III, we
assume AP configuration as the initial state. As the current is
increased, micromagnetic simulations give a fast red-shifting
branch corresponding to $\rm IPP_{-}$ (precession around the $-x$
axis) -- marked as 1 in Fig.~\ref{Fig:AP_both} (open squares). Fast
amplitude increase with increasing current soon leads to switching
towards the P state and damped oscillations stabilizing P
state are observed (range 2). However, as certain threshold
current is reached (P is now the initial state), the STT (region
I) counterbalances the damping and the second red-shifting branch,
$\rm IPP_{+}$, is observed (range 3). Micromagnetically, the
only difference between starting with initial P or AP state (open circles and open squares respectively) is the
appearance of first fast red-shifting branch.

Analogous comparison (see full circles for initial P state and full squares for initial AP states, Fig.~\ref{Fig:AP_both}) in the macrospin approximation leads to
similar conclusions, but now an additional mode is visible. When
current is high enough and assuming the AP state as the simulation
is initialized, the system might be forced to move into region II
leading to the appearance of OPP marked as 4 (full squares). Earlier in this
section it was shown that after crossing from region I through
the critical angle into region II, the static state was
observed. Here, however, dynamics in region II is forced by
the initial configuration and therefore the OPP can be observed.
In other words, in the single domain approximation the region I
supports $\rm IPP_{+}$, region II supports static state or OPP
(depending on the preceeding configuration), whereas in region
III the $\rm IPP_{-}$ mode can be observed. This result is
consistent with the one reported in [\onlinecite{Gmitra2006:APL}].

\subsection{Etched geometry}

In the extended structures (Fig.~\ref{Fig:extended}a) as discussed above, the interlayer coupling field (ICF) could be neglected, and the initial self-magnetostatic field together with the STT determined the induced dynamics
(region I, II or III). However in the etched geometry, schematically shown in Fig.~\ref{Fig:extended}b, the ICF can no longer be neglected. We have calculated this field micromagnetically, and by neglecting large OP edge
values we have estimated -36mT IP as an average ICF. This
is a significant contribution and therefore dynamics different
from that obtained for extended structures is expected.

\subsubsection{Initial P state}

\begin{figure}[!t]
\centering
\includegraphics[angle=0,width=1\columnwidth]{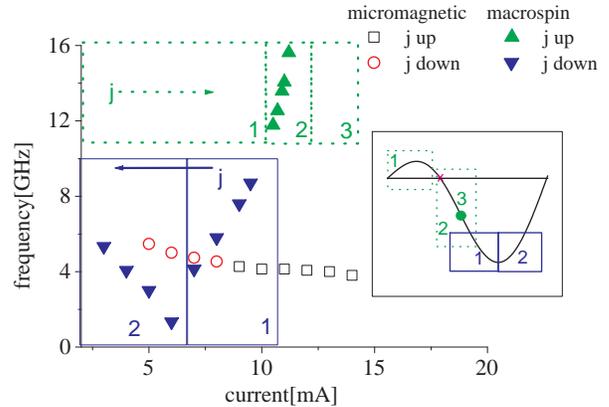}
\caption{The influence of current increase and decrease
         in the etched pillar geometry on the dynamical response of the system
         with P initial state. The corresponding torque diagram is shown in the inset.
         Note asymmetry in the macrospin frequency response -- OPP with current increase
         and transition from OPP to IPP with current decrease are observed.
         This hysteretic behavior originates from the asymmetric torque angular
         variation and the existence of the critical angle making dynamic transition
         between region I and II prohibited.}
\label{Fig:P_sweep}
\end{figure}

The dynamical response reveals now some new interesting features.
The macrospin approximation with the initial P state leads to the
steady OPP, which appears at 10.4 mA when the current is increased,
as indicated in Fig.~\ref{Fig:P_sweep} (corresponding range for observed mode marked with dotted lines). As the threshold current
for OPP is reached, the  blue-shifting branch appears in range
2. This OPP is not preceded by any IPP oscillations -- range
1, because the additional contribution from ICF places the system directly in region II. As the frequency of OPP oscillations increases with increasing
current, the corresponding amplitude decreases and the angle
between the magnetic moments of both layers approaches the
critical angle. When this angle is reached, the static state
discussed in the preceding subsection is observed (point 3). Now
we start to decrease current. The system is initially in the
static point and dynamic range is marked with solid lines in Fig.~\ref{Fig:P_sweep}. The OPP appears then in the range 1, and the amplitude and $\theta$ increase as the current decreases. The torque minimum is then passed and the
system moves to the region III, which results in $\rm IPP_{-}$
mode in range 2, where the amplitude decrease (with decreasing current) results effectively in the frequency redshift with current. The asymmetry in the
macrospin frequency response to increasing and decreasing current
is clearly a consequence of the torque asymmetry, the existence of
a critical angle, and irreversibility of the transition from region I to region II.

Dynamics in the micromagnetic model is simpler. Due to the effect
of ICF, the system directly switches to the region III, and only
one red-shifting branch $\rm IPP_{-}$ is observed. This mode is
qualitatively equivalent to the macrospin one in range 2 observed
for decreasing current.

\subsubsection{Initial AP state}

\begin{figure}[!t]
\includegraphics[angle=0,width=1\columnwidth]{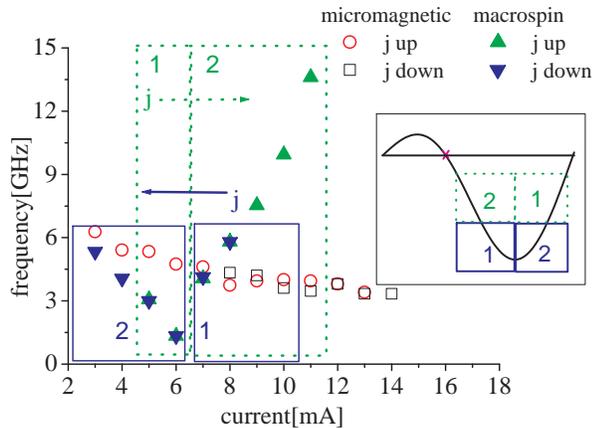}
\caption{The same as in Fig.~\ref{Fig:P_sweep}, but for the AP initial state. Now
no frequency hysteretic behavior with current is observed in the
macrospin model, and the dynamics evolves reversibly between the
regions II and III (inset). Micromagnetically only the IPP mode in
the region III is observed.} \label{Fig:AP_sweep}
\end{figure}

As discussed above, starting from the initial P state introduced
in the macrospin approximation hysteretic dependence in
frequency vs. current response. Clearly it is due to the
asymmetric shape of the torque angular dependence. This effect,
however, disappears when the initial state is AP. When the current
is increased (dotted in Fig.~\ref{Fig:AP_sweep}), the system is directly placed in region III supporting red-shifting branch
($\rm IPP_{-}$) in range 1 (inset Fig.~\ref{Fig:AP_sweep}). As the amplitude increases with increasing current, transition to range 2, where OPP
oscillations are triggered, is observed. In the torque diagram this is
equivalent to the transition from region III over the torque
minimum to region II. Since the critical angle is not crossed,
this transition stays reversible and no hysteretic behavior in the
frequency response is observed.

As before, micromagnetically no OPP was found. This indicates that
the system supports stable oscillations only in region III.

\subsection{Influence of the exchange field}
\subsubsection{The extended geometry}
\begin{center}
  \begin{figure}[h]
    \includegraphics[angle=0,width=0.8\columnwidth]{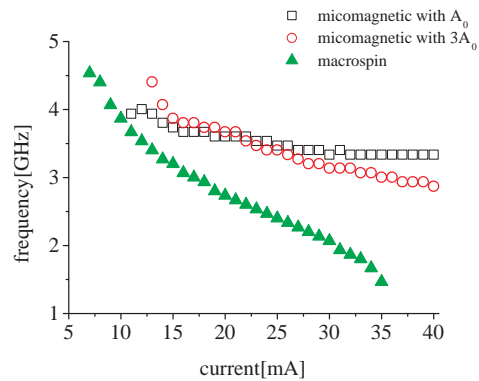}
    \caption{ The dynamic response of the system for two different
            values of the exchange constant, compared to the macrospin
            results. Micromagnetically faster redshift is observed for
            larger exchange constant as it favors more homogenous magnetization
             and accelerates the dynamics.}
  \label{Fig:exchange}
  \end{figure}
\end{center}

An open question is why the OPP modes obtained in the macrospin
model and also reported experimentally\cite{Boulle2008:PRB}, have
not been found in micromagnetic simulations. As has been shown,
to observe the OPP-associated blueshift, one has to force the
system dynamics in region II. Moreover, we have learnt that
the appearance of the macrospin 'spin-up' state was a consequence
of system crossing over the critical angle. This has not been
reached micromagnetically due to the inhomogenous nature of the
magnetization (finite exchange field) in the model. Therefore, one
should expect that the appearance of OPP in micromagentic model is
hindered by the underestimated exchange field, and that increase of the exchange constant should lead to a convergence of
both models. Using the bulk exchange constant for thin films might cause a great underestimation of exchange fields in these structures. Larger values of exchange
constants, as compared to the standard bulk ones, have been reported in Py
dots\cite{Lai:JMMM} and thin films \cite{Scholl:PRB}. Therefore,
we have investigated magnetization dynamics in the extended
geometry for the following values of the exchange constant:
$0.75A_{0}$, $2.5A_{0}$, $3A_{0}$, $4A_{0}$, $10A_{0}$. The
frequency-current behavior with $A= 3A_{0}$ is compared to the
results of macrospin model in Fig.~\ref{Fig:exchange}. Clearly
increasing the exchange constant changes the slope of
micromagnetic frequency redshift towards macrospin results. We
conclude that the finite exchange energy favoring inhomogenous
magnetization state causes this slope difference retarding the
dynamics.

\subsubsection{Etched geometry}

\begin{figure}[!t]
  \subfigure[]{\includegraphics[width=0.48\columnwidth]{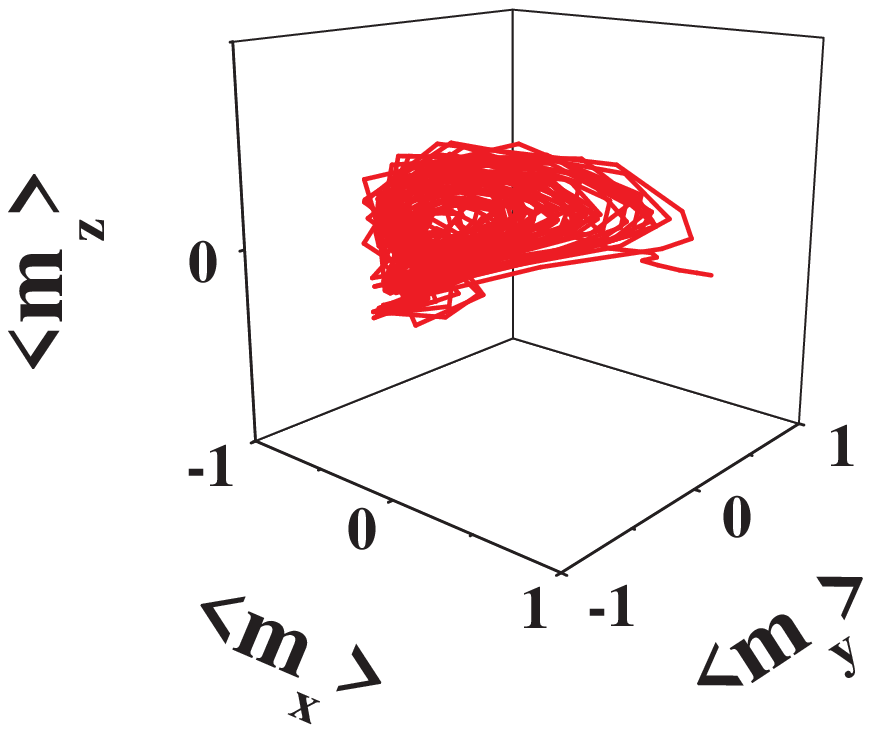}}
  \hfil
  \subfigure[]{\includegraphics[width=0.48\columnwidth]{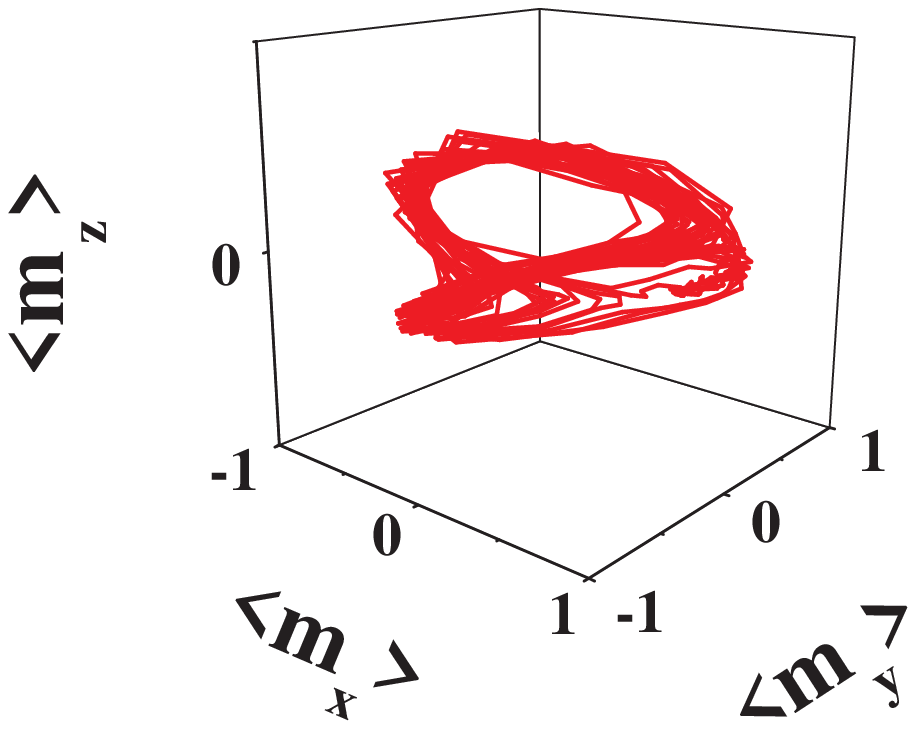}}
  \caption{Visualization of the exchange constant influence on the
          magnetization dynamics in the etched geometry at I=13 mA.
          Magnetization orbit for (a) $\rm A=A_{0}$ typical for region III, and (b) $\rm A=2.5A_{0}$ approaching the border
          between regions III and II}
  \label{Fig:orbit_exchange}
\end{figure}

So far in this geometry (ICF included) we have found that as the
current was increased micromagnetically induced dynamics in region III imposed frequency redshift and the macrospin dynamics in region II supported OPP-associated blueshift. Still we want to check whether micromagnetic dynamics can be shifted to region II by increasing the exchange constant. The micromagnetic
temporal evolution of the averaged magnetization component at I=13
mA for $A= A_{0}$ and $A= 2.5A_{0}$ results in different orbits,
Fig.~\ref{Fig:orbit_exchange}a and Fig.~\ref{Fig:orbit_exchange}b,
respectively. Clearly in the first case the ICF places the system
in region III forcing $\rm IPP_{-}$ dynamics. However, as the
exchange field is increased, which favors uniform magnetization,
an open clamshell orbit is formed (Fig.~\ref{Fig:orbit_exchange}b)
shifting the dynamics towards the border between region III
and II. One should note that the cross-over between the regions is
impossible in this geometry since the ICF has a significant
contribution to the effective field and hinders the appearance of
OPP.

\subsection{Comparison to experimental data}
\begin{center}
  \begin{figure}[h]
    \includegraphics[angle=0,width=0.9\columnwidth]{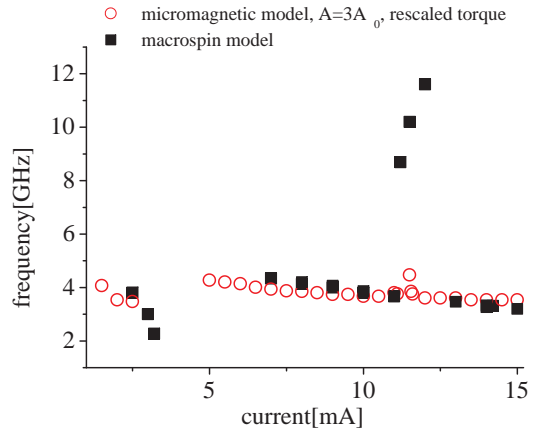}
    \caption{Comparison of frequency vs. current behavior at 0mT between macrospin
             model and micromagnetic model with $\rm A=3A_{0}$.
             As described in the text, in the micromagnetic model the torque strength
             has been scaled by a factor of 0.5 in order to counterbalance the
             effect of its inhomogeneous nature.
             The threshold for OPP predicted here by both models coincides with the
             experimental one [\onlinecite{Boulle2008:PRB}].
             The magnitude of the frequency jump associated with the transition IPP-OPP
             in the macrospin approximation does not fit the experimental
             values and micromagnetic approach proves to be more accurate.}
  \label{Fig:rescaled}
  \end{figure}
\end{center}
Still no OPP (supported in the dynamic region II) has been
predicted micromagnetically, although such modes were reported
experimentally\cite{Boulle2008:PRB} in extended structure. However
previous paragraphs have given some important clues. We have
learnt about the importance of the underestimation of the exchange
field. Therefore, we assume $A=3A_{0}$ for further study.
Secondly, as the magnetization always stays inhomogeneous to some extend,
the toruqe calculated locally (cell by cell) inherits this
inhomogeneity and we shall scale the torque strength by a factor
of 0.5 to counterbalance this effect. Thirdly, micromagnetically
the transition from dynamic region I to II was impossible. Thus in
order to observe OPP one has to force dynamics in region II by
forcing the transition from region III to II (i.e. imposing AP
initial state). As presented in Fig.~\ref{Fig:rescaled}, indeed
under all above mentioned assumptions both models converge.
Micromagnetic dynamics is forced first in region III
supporting $\rm IPP_{-}$, then switching towards P state takes
place (range of current where no sustained oscillations are observed, as
discussed before), and then $\rm IPP_{+}$ branch (region I) is
triggered. At a certain threshold, however, the dynamics in
region II supporting blueshift can be obtained. The threshold of
this OPP coincides with the experimental one from
[\onlinecite{Boulle2008:PRB}]. The magnitude of the frequency jump
associated with the transition IPP-OPP in the macrospin
approximation does not fit to the experimental values, and
micromagnetic approach proves to be more accurate. The fact that
experimentally OPP was reported by starting from P state (opposite
to our results) means that transition between region I and II
prohibited micromagnetically is experimentally possible due to
thermal activation. As our simulations neglect the effect of
thermal fluctuations, the dynamic region II can be only reached by
transition from region III.
   \begin{figure}[h]
    \includegraphics[angle=0,width=1\columnwidth]{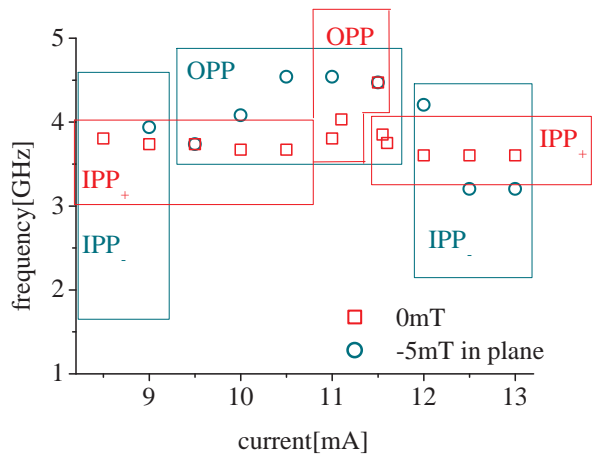}
    \caption{Results of micromagnetic model at zero and low applied field.
             The dynamic response at 0mT and -5mT applied in-plane.
             In absence of external field $\rm IPP_{+}$ transition into OPP at 10.5mA
             and back to $\rm IPP_{+}$ above 11.5mA is observed.
             At -5mT the transitions from $\rm IPP_{-}$ to OPP and back to $\rm IPP_{-}$ are observed at 9.5mA and 12mA respectively.
             Both, blueshift linear at 0mT and nonlinear at -5mT predicted
             micromagnetically, are consistent with the experimental results
             [\onlinecite{Boulle2008:PRB}].
             Macrospin ceases to show the saturation regime observed experimentally
             at -5mT but full micromagnetic study correctly predict this feature.}
 \label{Fig:micro_low_field}
 \end{figure}
One should note that not all regimes (dynamics in all regions I,
II, III) predicted by the simulations for the extended geometry were
observed experimentally. Low angle IPP does not provide enough
output power to be measured via GMR effect. Therefore in order to
conduct meaningful comparison we have concentrated on the OPP
regime (region II), which was both predicted numerically and
observed experimentally. In the absence of external field
satisfactory qualitative agreement has been reached
(Fig.~\ref{Fig:micro_low_field}, compare to Fig. 6a in Ref.
[\onlinecite{Boulle2008:PRB}]). Not only the threshold current
($I_{th,sim}=11$ mA compared to $I_{th,exp}=10$ mA) but also the
agility (0.6 GHz/mA and 0.7 GHz/mA respectively) are in good
agreement. The remaining quantitative difference in frequency
values is a consequence of uncertain estimation of the factors
entering micromagnetic model, like saturation magnetization and/or
damping. Moreover, the dynamics at -5mT IP field reveals that OPP
threshold current is smaller with respect to 0mT case, which is
again consistent with the experimental results. Interestingly,
both approximately linear at 0mT and nonlinear at -5mT behavior of
the frequency as a function of current, are well reproduced in
frames of micromagnetic model. Note that this feature was not
reported in single domain model. Furthermore, micromagnetic model
predicts experimentally observed saturation at -5mT, i.e. in a
certain current range the frequency stays relatively constant. In
the model it is associated with the large angle orbit
stabilization around the torque minimum, i.e. the system
approaches the border of dynamic regions II and III, and the
torque shape becomes flat around its minimum. Furthermore, current
increase leads to transition between OPP (region II) and $\rm
IPP_{-}$ (region III) and reappearance of the clamshell orbit.
However, since neither was the IPP reported in the experiment
prior to the appearance of OPP, as predicted micromagnetically,
nor it could have been detected following OPP regime (because of
low output power) so obviously the experimental cut off current
refers to the threshold current for IPP reappearance in the model.
Note that micromagnetically the main mode (supported over largest
range of currents) in case of 0mT was the $\rm IPP_{+}$, whereas
even low applied field of -5mT IP forced the dynamics in region
III and therefore the $\rm IPP_{-}$ was observed as the main mode.
In other words in the absence of external field the increased
exchange constant enabled direct dynamics in region II (and
associated linear frequency vs. current slope) and additional
field forced the transition from region III to region II resulting
in the appearance of the saturation regime. Clearly the OPP in
both cases is preceded by different dynamics.
\begin{center}

\end{center}

\section{Conclusions}\label{Sec:conclusion}
\typeout{--------------------------------------}
\typeout{Conclusion}
\typeout{--------------------------------------} The results
emerging from the macrospin and micromagnetic model very often do
not converge. Lack of this convergence is attributed to a couple
of factors. It was shown that the discrepancies arise from the
inhomogenous magnetization state and finite exchange field. We
concluded that the OPP in the macrospin model might appear
as a consequence of the 'spin-up' static state state,
which seems to be characteristic of the macrospin model. On the
other hand, the absence of OPP in the micromagnetic model was
identified as a combined consequence of the underestimation of the
exchange constant and the role of the initial self-magnetostatic
field. By setting the initial state and the exchange constant
favoring the appearance of the OPP, a good qualitative agreement
was reached between the predictions of both models. Thus, only
micromagnetic model has proven to predict correctly dynamics
reported experimentally.

\section*{Acknowledgment}
This work was supported by EU Training Network SPINSWITCH
(MRTN-CT-2006-035327). JB also acknowledges support by funds from
the Ministry of Science and Higher Education as a research project
in years 2006-2009 within the EUROCORES Programme FoNE (project
SPINTRA).

\bibliographystyle{unsrtg}
\end{document}